\documentclass[aps,superscriptaddress,eqsecnum,nofootinbib,preprintnumbers]{revtex4}
\usepackage{amsmath}
\usepackage{amssymb}
\usepackage{amsfonts,amsthm,bm}
\usepackage{color,xcolor}
\usepackage[greek,english]{babel}
\usepackage{titlesec}
\usepackage{graphicx,epsfig}
\graphicspath{{figures/}{./}}
\usepackage{float}
\usepackage{subfigure}
\usepackage{tikz}
\newcommand{\m}{\mu}
\newcommand{\n}{\nu}

\newcommand{\g}{\gamma}

\newcommand{\be}{\begin{eqnarray}}
	\newcommand{\ee}{\end{eqnarray}}
\newcommand{\bea}{\begin{eqnarray}}
	
	\newcommand{\eea}{\end{eqnarray}}

\def\a{\alpha}
\def\b{\beta}
\def\g{\gamma}

\def\k{\kappa}
\def\m{\mu}
\def\n{\nu}

\def\th{\theta}

\def\Ref{\ref}

\newcommand{\beq}{\begin{equation}}
\newcommand{\eeq}{\end{equation}}
\newcommand{\bseq}{\begin{subequations}}
	\newcommand{\eseq}{\end{subequations}}

\begin{document}

\leftline{KCL-PH-TH/2022-{\bf 37}}
%\vspace{0.1cm}

\title{Axion induced angular momentum reversal in Kerr-like black holes}

\author{Nikos Chatzifotis }
\email{chatzifotisn@gmail.com} \affiliation{Physics Department, School of Applied Mathematical and Physical Sciences,
		National Technical University of Athens, 15780 Zografou Campus,
		Athens, Greece.}

\author{Panos Dorlis}
\email{psdorlis0@gmail.com}
\affiliation{Physics Department, School of Applied Mathematical and Physical Sciences,
		National Technical University of Athens, 15780 Zografou Campus,
		Athens, Greece.}

\author{Nick E. Mavromatos}
\email{mavroman@mail.ntua.gr}
\affiliation{Physics Department, School of Applied Mathematical and Physical Sciences,
		National Technical University of Athens, 15780 Zografou Campus,
		Athens, Greece.}
\affiliation{Particle Physics and Cosmology Group, Department of Physics, King's College London, Strand, London WC2R 2LS, UK.}

\author{Eleftherios Papantonopoulos}
\email{lpapa@central.ntua.gr} \affiliation{Physics Department, School of Applied Mathematical and Physical Sciences,
		National Technical University of Athens, 15780 Zografou Campus,
		Athens, Greece.}
\vspace{1cm}

\begin{abstract}
We consider a pseudoscalar axion-like field coupled to a Chern-Simons gravitational anomaly term. The axion field backreacts on a rotating Kerr black hole background, resulting in modifications in the spacetime. In an attempt to determine potentially observable signatures, we study the angular momentum of the system of the modified Kerr-like black hole and the axionic matter outside the horizon of the  black hole. As the strength of the coupling of the axion field to the Chern-Simons term is increasing, the requirement that the total angular momentum of the system  remain constant forces the black hole angular momentum to decrease. There exists a critical value of this coupling beyond which the black hole starts to rotate in the opposite direction, with an increasing magnitude of its angular momentum. We interpret this effect as a consequence of the exchange of energy  between the axionic matter  and the gravitational anomaly, which is sourced by the rotating black hole.
\end{abstract}

\maketitle

%---------------------------------

\section{Introduction}

The behaviour of a black hole interacting with the surrounding matter can be studied quite accurately
by means of perturbation techniques, where the dynamics of matter, electromagnetic and
gravitational waves is taking place in a fixed background spacetime generated by the black hole.
In the case that the background metric is that of a static spherically symmetric Schwarzschild black hole,
orbits of surrounding particles can be described easily and also electromagnetic  and gravitational
perturbations are well studied.

{However, the black holes in a realistic Universe, are rotating, which, when idealised, are described by the Kerr metric~\cite{kerr}.
In the case of a rotating black hole described by the Kerr metric, the description of the behaviour of matter
is more complicated than in the non-rotating case. The metric in this case is  only stationary,  whilst the spherical symmetry has been replaced by axisymmetry. A further difficulty is the complexity of the coordinate systems for describing processes
near a Kerr black  hole.

From a theoretical point of view, Kerr-like black hole solutions appear in a plethora of modified theories of gravity. Of particular interest to us in this work are the Chern-Simons (CS)-gravity theories~\cite{Jackiw, Zanelli:2005sa,Alexander,yunes1,yunes2}, which contain
second-order curvature corrections (fourth-order in derivatives), and also characterise the low-energy-limit of  string theory~\cite{kerr1,kerr2,kerr3}. The CS modification is a total
derivative, and thus innocuous unless pseudoscalar (axion-like) fields couple to it, in which case it deforms non-trivially the gravitational field of spinning black holes. Such deformations are described by a modified Kerr geometry. In the context of string theory  the axion fields are part of the excitation spectrum, and they are of two types~\cite{witten}. One, is the so-called string-model independent axion, which in (3+1)-dimensions is dual~\cite{kerr3,witten} to the field strength of the so-called Kalb-Ramond, spin-one, antisymmetric-tensor field,
which plays the r\^ole of a totally antisymmetric torsion in the geometry. The second type of axions in string theory arise from compactification of the extra dimensions.

Rotating black holes have been studied extensively in such CS theories. First, they were found as solutions  of second-order field equations for the metric stemming from string theories, in the presence of axions, in~\cite{kerr1,kerr2,kerr3}, and more recently in  \cite{CSblackhole,Gonzalez:2010vv}. Potential signatures of CS theories in the emission of gravitational waves,  produced in the inspiral of stellar compact objects into massive black holes, both for intermediate and extreme mass ratios, have been investigated in \cite{Sopuerta:2009iy}. Also in \cite{Canizares:2012is}, the authors attempted to answer the question as to what extent the extreme mass ratio inspirals observations with a space based gravitational wave observatory like LISA  might be able to distinguish between CS theories and General Relativity (GR).

In the generic CS gravity theory,  slowly rotating black hole solutions, in the presence of axion fields  back reacting on the geometry, have been recently investigated in  \cite{Chatzifotis:2022mob}, extending non-trivially previous results~\cite{kerr1,kerr2,kerr3,yunes1,yunes2}, \cite{Myung:2020etf,Myung:2021fzo,Zou:2021ybk}. In \cite{Chatzifotis:2022mob}, the back reaction of the axionic field on the slowly-rotating background metric has been studied to  arbitrary order  in perturbation theory of an appropriate parameter, $\gamma$, proportional to the square of the axion-CS interaction coupling. Such an expansion allows one to go arbitrarily close to the horizon of the hairy (with axion-hair~\cite{kerr1,kerr2,kerr3}) black hole. One can thus study the behaviour of the axionic matter outside the horizon of the slowly-rotating Kerr-type metric and see how this matter distribution influences the Kerr-like black hole itself.

The behaviour of matter in the neighbourhood of rotating black holes has been extensively studied by now.
In their pioneering work, Bardeen, Press and Teukolsky  \cite{Bardeen:1972fi}, have studied the physical processes around the black holes, together with their properties and interactions with the surrounding matter.
To avoid the complexity in the perturbation analysis,
in most cases it is assumed that the background metric is slowly rotating. The method used in \cite{Bardeen:1972fi} has essentially assumed that the non-rotating matter  cancels out the “frame-dragging” effects of the black hole rotation. In a sense, such a matter is co-rotating with the black hole in such a way that the physical processes can be analyzed in their own frame.  The authors of  \cite{Bardeen:1972fi} have used the method of locally non-rotating frames, which gives a
clear picture of the process of energy extraction. The amount of such energy, however, for realistic black holes is not  astrophysically significant.

In recent studies~\cite{Lukes-Gerakopoulos:2010ipp} the extreme mass-ratio inspirals of a small-mass compact object
orbiting around a  massive compact object, that is described by a  Kerr-like black hole, have been studied with the aim of detecting
possible gravitational wave signals. The geodesic motion in a spacetime deviating from the Kerr metric was studied in \cite{Zelenka:2017aqn}.
It was shown that the case exhibits chaotic behavior. In \cite{Destounis:2021mqv}, it was demonstrated  that in the case of extreme-mass-ratio inspirals of a compact rotating object, the frequencies of the gravitational waves increase abruptly. Moreover, the question was raised as to whether this behaviour had a genuine astrophysical origin associated with fundamental spacetime symmetries associated with supermassive compact objects. The authors of \cite{Destounis:2020kss}, examined the question whether the detection of gravitational waves from extreme mass ratio inspirals would place strong constraints on hypothetical departures from a Kerr description for astrophysically stable black holes. It was further found in \cite{Gonzalez:2018lfs}, by means of studying the  motion of particles in the background of  rotating linear dilaton black holes, that these black holes  act as a particle accelerator.

An extensive study of the behaviour of angular momentum and  the geodesics of the  Kerr family of solutions of Einstein and Einstein-Maxwell systems had been carried out in \cite{carter}. In the case of the Kerr black hole, the total angular momentum is given by $Ma$, where $M$ is the mass  and $a$  the angular parameter of the rotating black hole.
In this respect we stress that in \cite{Chatzifotis:2022mob},  the presence of the axion field coupled to the CS term, which  back reacts  on the Kerr Black-hole background,  implies that the total angular momentum consists of the contributions of both the axion field and the hairy Kerr-like black hole itself, which arises as a consequence of the axionic back reaction on the geometry.

In this work we extent the study in \cite{Chatzifotis:2022mob} on the axion-modified-Kerr black hole system  by searching for potentially observable effects associated with bound trajectories of satellites in such backgrounds.
 As in \cite{Chatzifotis:2022mob}, a key r\^ole in the analysis will be played by the aforementioned perturbative parameter $\gamma$, which is proportional to the coupling of the axion field with the CS term and inversely proportional to the square of the total energy of the spacetime. We stress that it is this dimensionless parameter, rather than the coupling constant itself, that appears in the correction function of the modified Kerr-like black hole solution and measures the strength of the backreaction on the geometry. The axion matter outside the horizon of the black hole acquires an angular momentum in such a way that the total angular momentum of the system (black hole and axionic matter) remains constant and equal to $Ma$.
We shall demonstrate that as $\gamma$  increases,  the black hole angular momentum decreases and there exist a critical value $\gamma_c$ beyond which the black hole angular momentum increases while the black hole starts to rotate to the opposite direction. As such, the angular momentum of the black hole horizon may in principle reach large values in magnitude, while the slowly rotating approximation remains valid with a total angular momentum equal to $Ma$.  We understand this effect as a counterplay of two competing systems; the Kerr-like black hole and the axionic matter rotating outside the horizon of the black hole.

With the CS coupling of the axion field, the gravitational equations of motion are modified in such a way that only the effective energy-momentum tensor is conserved, implying an exchange of energy between the axion field and gravity \cite{Mavromatos:2021urx}. This interplay between the axion and the gravitational field is responsible for highly rotating effects around the black hole, albeit, in a slowly rotating spacetime. We present such effects by studying particles behaviour around the black hole. The equatorial motion of particles arround the Kerr black hole have been extensively studied in \cite{Pugliese:2011xn}. In the case of the extreme Kerr, a qualitatively different structure of the effective potential for the co-rotating and counter-rotating orbits distinguishes these orbits and since the extreme Kerr corresponds to a highly rotating black hole, this can be understood as a highly rotating effect.

 A similar distinction between co-rotating and counter-rotating particles is observed in our case, too. The crucial difference, however, is that our spacetime is slowly rotating. In \cite{Pugliese:2011xn} there is no interaction of gravity with any matter field and as such a highly rotating spacetime corresponds to a highly rotating black hole and vice versa. However, this is not the case if non-trivial interaction of a matter field with gravity is introduced. In other words, such a non-trivial interaction can construct a spacetime  which is slowly rotating as a whole, but with an internal structure in which a highly rotating event horizon is allowed, leading to highly rotating effects.

The work is organized as follows: in Section \ref{sec2}, we review the particle trajectories around a slowly-rotating Kerr black hole. In Section \ref{sec3}, we study the particle trajectories around the slowly rotating  Kerr-like black hole, arising from axion  back reaction on the geometry. In Section \ref{sec4}, we calculate both, the black hole and axion-matter angular momentum and find a counter-rotating behaviour for the angular momentum of the black hole. Finally, in Section \ref{sec5} we give our conclusions.

\section{Geodesics in a Slowly Rotating Kerr Black Hole}
\label{sec2}

In this section we will review the motion of a particle around a Kerr black hole slowly rotating about, say, the $z$ axis of a Cartesian coordinate system,
which is described by the line element:\footnote{In this article we denote $M=\mathcal{M}G$, where $\mathcal{M}$ is the black hole mass with the correct units.}
\begin{equation}
	ds^2= -\left( 1-\frac{2M}{r} \right)dt^2+\frac{dr^2}{\left( 1-\frac{2M}{r} \right)}+r^2d\Omega^2+2g_{t\phi}(r,\th)dtd\phi~.
	\label{eq:metricsolution1}
\end{equation}
where
\begin{equation}
g_{t\phi}(r,\th)=-\frac{2Ma}{r}\sin^2(\th)~,
	\label{gtphi}
\end{equation}
The parameter $a$, which is assumed to be small in our slowly rotating approximation scheme, defines the angular momentum of the black hole per unit mass, as is derived from the corresponding Komar integral. As is well known, the metric (\Ref{eq:metricsolution1}) describes a stationary and axisymmetric spacetime, which possesses the Killing vectors
\begin{equation}
	k=\partial_t\;\;\text{and}\;\;\xi=\partial_\phi~,
	\label{kvec}
\end{equation}
where $k$ is timelike and $\xi$ is spacelike on the black hole  exterior and generates a 1-parameter group of isometries isomorphic to $U(1)$. According to these Killing vectors, there are three constants of motion, namely
\begin{equation}
	\label{Constantsofmotion}
	\begin{aligned}
		& E=-k^\m u_\m\rightarrow g_{tt}\dot{t}+g_{t\phi}\dot{\phi} = -E~,  \\
		& L_z= \xi^\m u_\m\rightarrow g_{t\phi}\dot{t}+g_{\phi\phi}\dot{\phi}=L_z~,   \\
		& g_{\m\n}u^\m u^\n=\epsilon~,
	\end{aligned}
\end{equation}
where $E$ is the particle energy, $L_z$ the $z$-component of its angular momentum, and
$u^\m$ is the tangent vector of the geodesic, with $\epsilon=-1,0$ for massive and massless particles, respectively.\footnote{Here, and in most of the article, we work in units of the particle (rest) mass, which is constant in our case, given that there are no external forces acting on the particle, which would affect its dynamics. Nonetheless, in what follows, when is necessary to express certain quantities in proper units, we will revert to using the symbol $\m$ for the particle mass.} Moreover, we assume an affinely parameterized geodesic, a fact that provides us with the third constant of motion. The first two constants of motion give
\begin{equation}
	\begin{aligned}
		& \dot{t}=\frac{g_{t\phi}L_z+g_{\phi\phi}E}{g_{t\phi}^2-g_{tt}g_{\phi\phi}}~,  \\
		& \dot{\phi}= -\frac{g_{t\phi}E+g_{tt}L_z}{g_{t\phi}^2-g_{tt}g_{\phi\phi}}~,
	\end{aligned}
\label{ELzconstants}
\end{equation}
while, substitution to the third one results in the equation
\begin{equation}
	\begin{aligned}
		&g_{rr}\dot{r}^2+g_{\th\th}\dot{\th}^2+V^E_{eff}(r,\th)=0~,\\
		\text{where}\;\;&V^E(r,\th)=\frac{L_z^2g_{tt}+E^2g_{\phi\phi}+2EL_z g_{t\phi}}{g_{tt}g_{\phi\phi}-g_{t\phi}^2}-\epsilon~.
	\end{aligned}
	\label{effectivepotentialaxisymmetric}
\end{equation}
The function $V^E(r,\th)$ depends on the particle energy $E$, which makes it not well defined as an effective potential. In order to define the effective potential, we  know  that when the kinetic energy vanishes, the total energy equals the potential one. Thus, by setting $\dot{r}=\dot{\th}=0$ and solving the equation $V^E(r,\th)=0$ with respect to $E$, the corresponding function can be interpreted as the effective potential. We may rewrite $V^E(r,\theta)$ as
\begin{equation}
	V^E(r,\th)=-\frac{g_{\phi\phi}E^2+2L_zg_{t\phi} E+L_z^2g_{tt}+\epsilon\tilde{\Delta} }{\tilde{\Delta}}~,
	\label{effectivePot}
\end{equation}
where $\tilde{\Delta}=g_{t\phi}^2-g_{tt}g_{\phi\phi}$. The latter corresponds to the metric determinant via $g=-g_{rr}g_{\phi\phi}\tilde{\Delta}$. Since the metric signature is $(-+++)$, its determinant is negative definite, implying $\tilde{\Delta}>0$. Setting now $V^E=0$, which naturally implies the vanishing of the numerator, we may extract a quadratic equation with respect to $E$, from which we obtain the  solution
\begin{equation}
	V^{(\pm)}(r,\th)=-\frac{L_z g_{t\phi}}{g_{\phi\phi}}\pm\frac{\sqrt{\tilde{\Delta}}}{g_{\phi\phi}}\sqrt{L_z^2-\epsilon g_{\phi\phi}}~.
\label{eq222}
\end{equation}
In terms of these, (\ref{effectivepotentialaxisymmetric}) becomes
\begin{equation}
	g_{rr}\dot{r}^2+g_{\th\th}\dot{\th}^2=\frac{(E-V^{(+)})(E-V^{(-)})}{\tilde{\Delta}}~.
	\label{energyconservationequation}
\end{equation}
The left hand side of the above equation is non-negative and vanishes at the turning points of the phase space at which $E=V^{(+)}$ or $E=V^{(-)}$. Thus, whether $E=V^{(+)}$ or $E=V^{(-)}$, depends strictly on the sign of $V^{(\pm)}$.

For the slowly rotating case that we are considering here, the absence of the ergosphere in the exterior region implies that the particles energy $E=-k^\m u_\m$ is positive definite. Focusing on the effective potential (\Ref{eq222}), we may distinguish the following two cases:
\begin{enumerate}
	\item $L_z g_{t\phi}<0$: This case corresponds to particles co-rotating with the black hole geometry.  In this case, $V^{(+)}>0$ for all the exterior of the black hole. Assuming $V^{(-)}<0$,  we find $-g_{tt}L_z^2>\epsilon\tilde{\Delta}$, which clearly holds in the stationary region, implying that $(E-V^{(-)})>0,\;\forall E$. Thus, the effective potential is given by $V^{(+)}$.
	\item  $L_z g_{t\phi}>0$: This case corresponds to counter rotating particles.	Obviously, $V^{(-)}<0$ and consequently $(E-V^{(-)})>0,\;\forall E$. Furthermore, assuming $V^{(+)}<0$, it follows: $-g_{tt}L_z^2<\tilde{\Delta}\epsilon$. Since, $\tilde{\Delta}>0$ outside the event horizon and $\epsilon=0,-1$, it follows that $V^{(+)}$ becomes negative, in the region determined by $g_{tt}>0$, i.e. in the interior region. Thus, $V^{(+)}$ can be interpreted as the effective potential in the sense that the allowed regions of motion are determined by $V^{(+)}\leq E$, since the right hand size of (\ref{energyconservationequation}) has to be non-negative and equal to zero at the turning points.
\end{enumerate}
We may therefore conclude that the effective potential in the stationary region is
\begin{equation}
	V_{eff}(r,\th)=-\frac{L_z g_{t\phi}}{g_{\phi\phi}}+\frac{\sqrt{\tilde{\Delta}}}{g_{\phi\phi}}\sqrt{L_z^2-\epsilon g_{\phi\phi}}~,
	\label{effectivepotentialstationary}
\end{equation}
in the sense that motion is allowed in regions where $E\geq V_{eff}(r,\th)$. Up to $\mathcal{O}(a)$, the above equation reads
\begin{equation}
	V_{eff}(r,\th)=\frac{\sqrt{r(r-2M)(L_z^2+r^2\sin^2(\th))}}{r^2\sin(\th)}+\frac{2 L_z M a }{r^3}+\mathcal{O}(a^2)~.
	\label{effectivepotentialKerrSolwlyRotating}
\end{equation}
which is plotted in the following figures on the equatorial plane.\footnote{The quantity $\m$ in the figures is the particle (rest) mass, in units of which we work throughout the article.}
\begin{figure}[h!]
	\centering
	\includegraphics[width=0.5\textwidth]{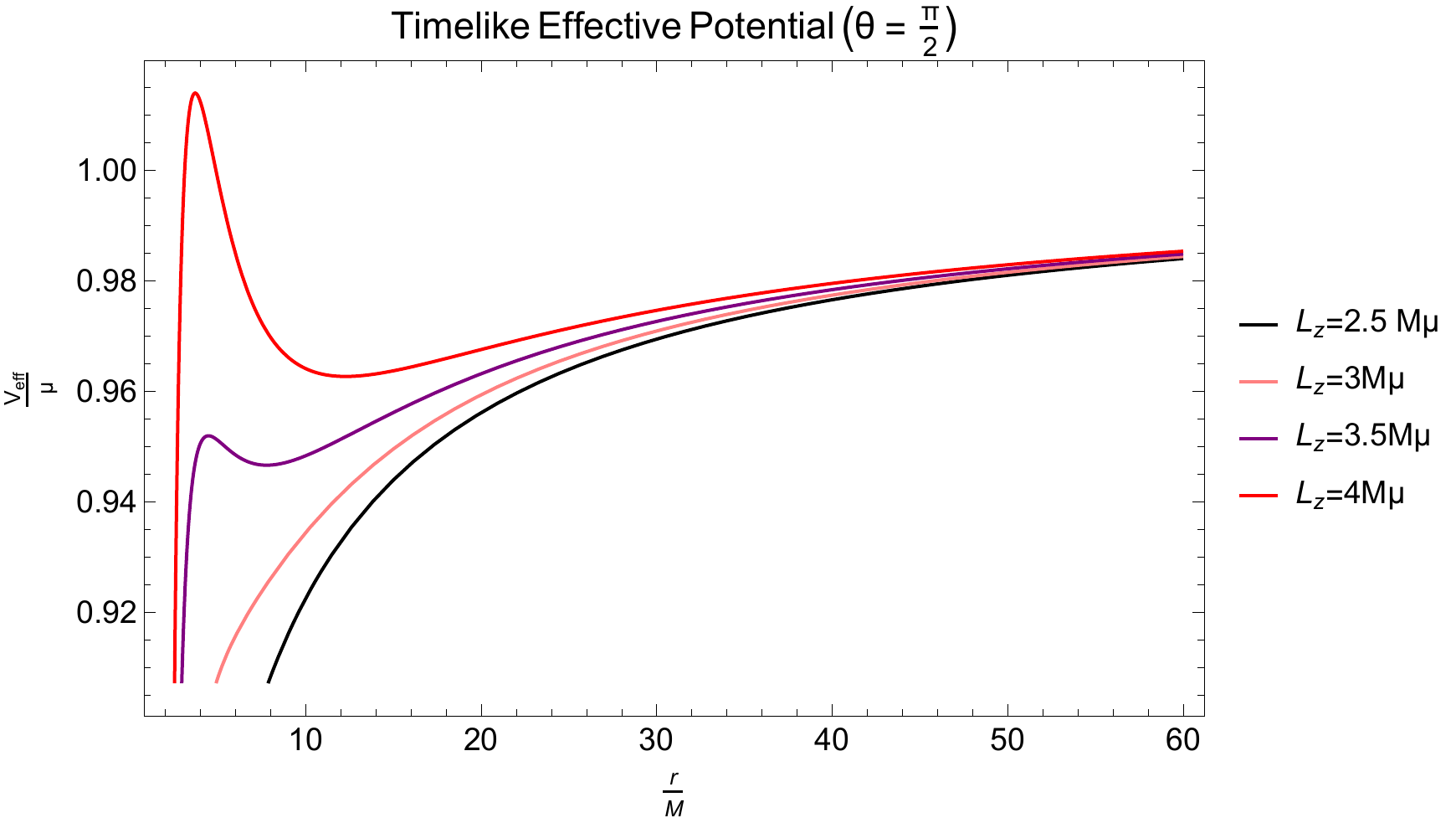}\hfil\includegraphics[width=0.5\textwidth]{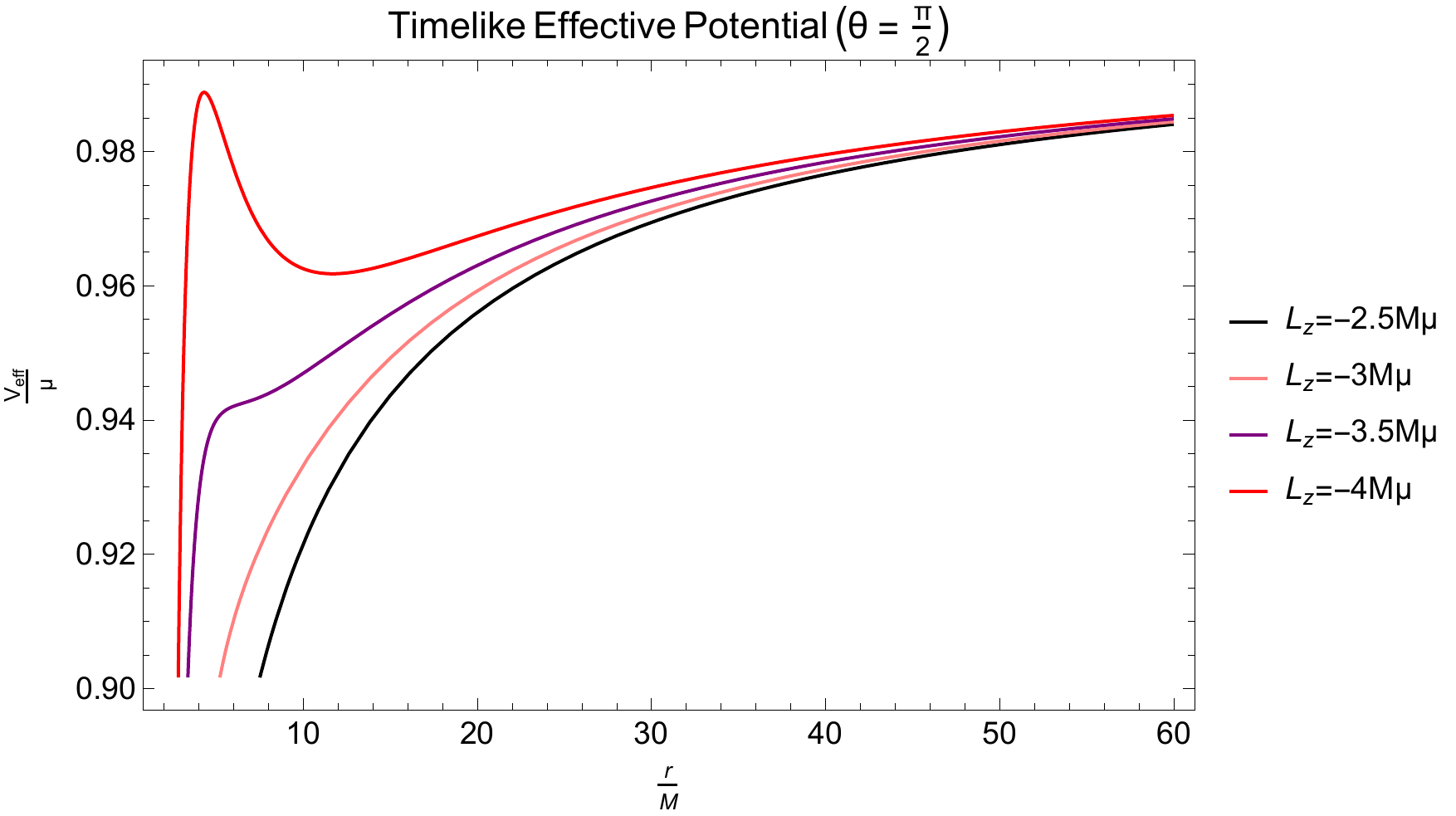}
		\caption{Effective potential at the equatorial plane for timelike geodesics for the slowly rotating Kerr metric. The parameter $a$ is fixed to $a=0.1$. .}
\label{fig1}
\end{figure}
\\

A local maximum and a minimum appear for increasing angular momentum, signifying for bound orbits.  However, spherical symmetry is absent and as such the potential is $\th$-dependent. This implies that the motion cannot be constrained on a single plane, without external forces. We can draw the turning points at the $\th-r$ plane by setting the effective potential equal to some energy, $V_{eff}=E$ and then plot the corresponding contour plot, as we do in Fig.~\ref{fig2}. By virtue of equation (\Ref{energyconservationequation}), the phase space points of $V_{eff}=E$ define curves of zero velocity (CZV), since the left hand side of the equation enforces $\dot{r}=\dot{\theta}=0$.

 \begin{figure}[h!]
	\centering
	\includegraphics[width=0.4\textwidth]{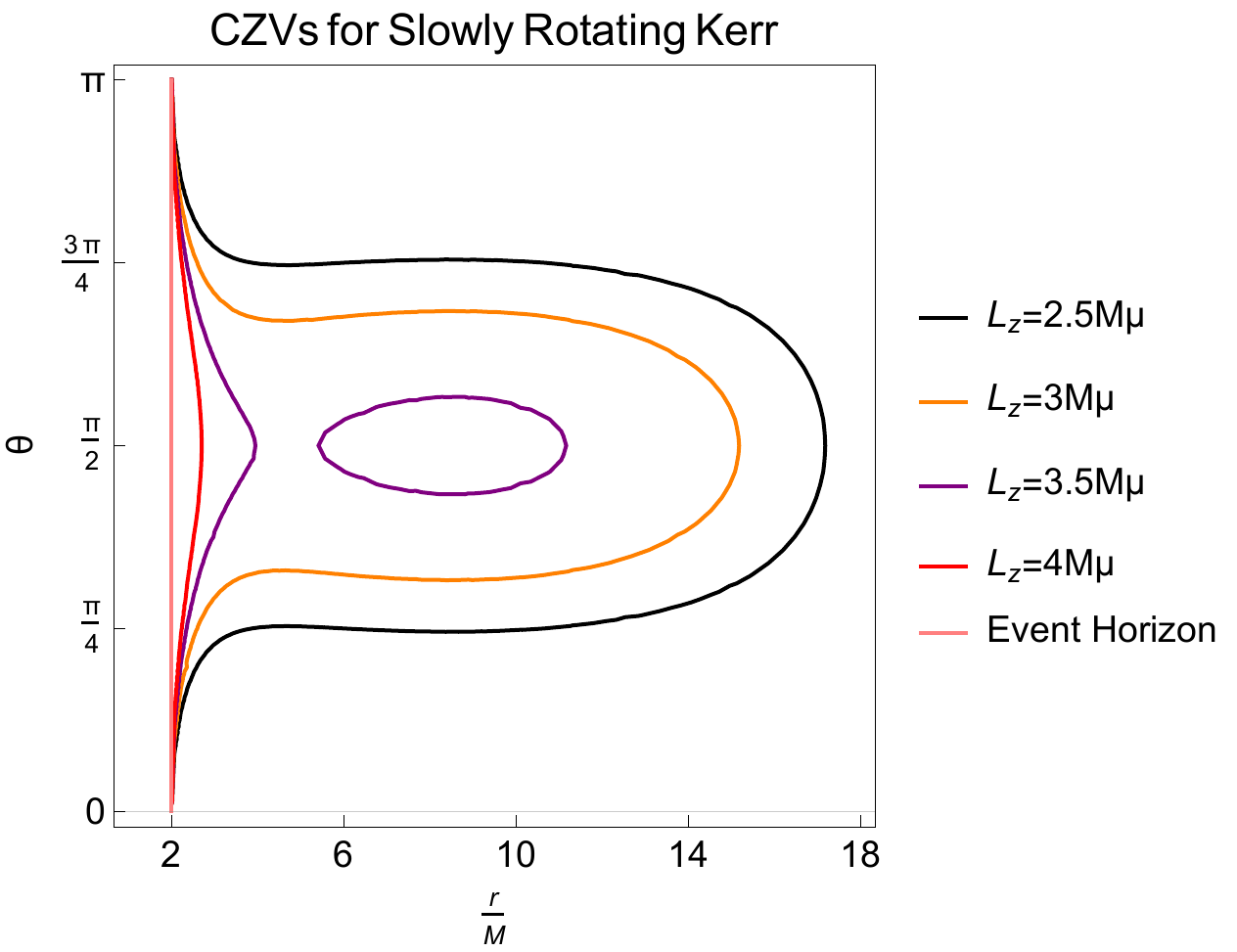}\hfil\includegraphics[width=0.4\textwidth]{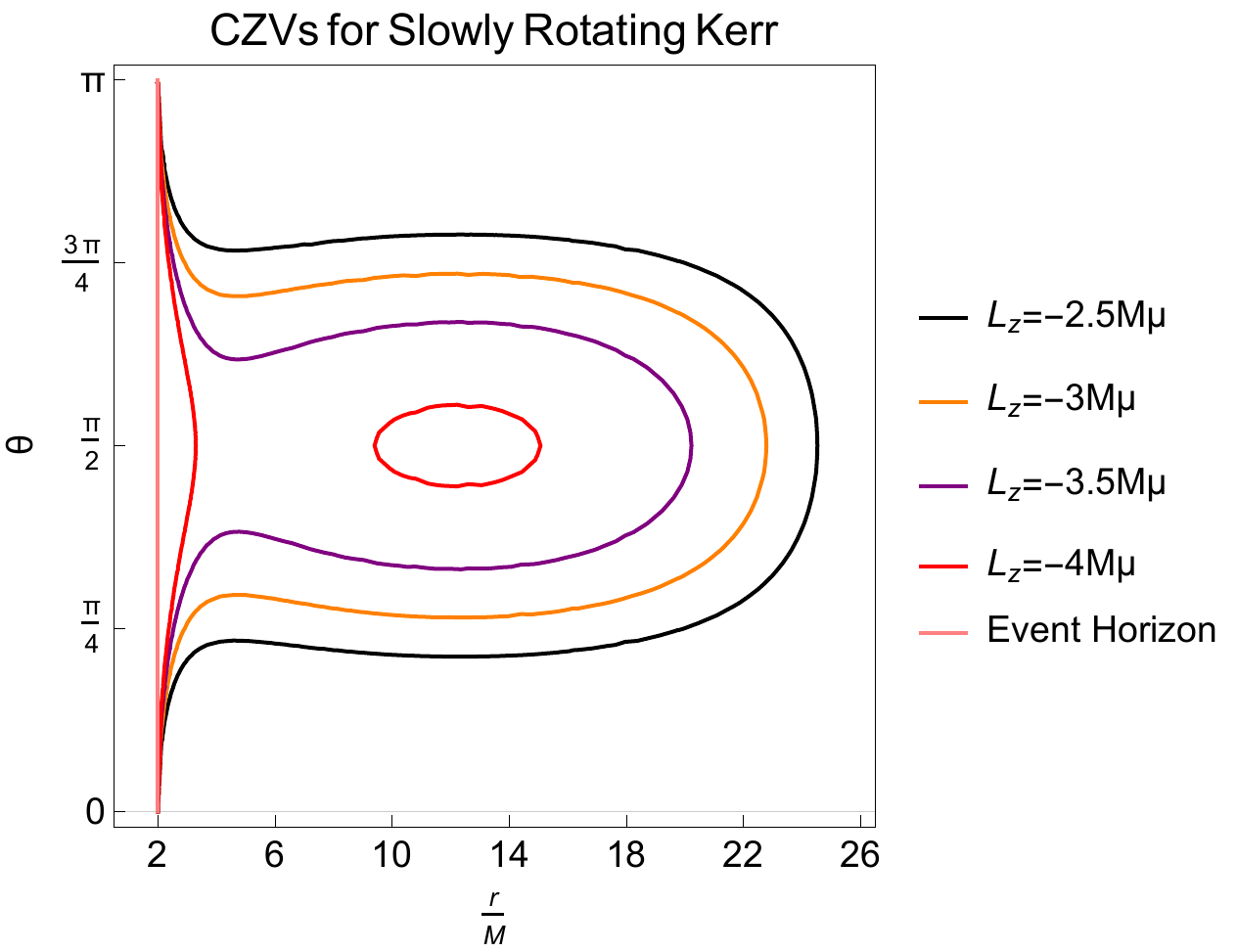}
	\caption{Curves of Zero Velocity (CZV) for timelike geodesics. The plots are evaluated for $M=1$, $a=0.1$ and $E=0.95$ (for the left hand site) and $E=0.9633$ (for the right hand site).}
	\label{fig2}
\end{figure}

 The graphs in the CZV figures were obtained for a particle's energy $E=0.95$ (in units of its rest mass $\m$). It is obvious that such particles cannot escape to infinity. We note that this does not mean that all of these orbits stay outside the event horizon. Whether a particle falls into the black hole or remains in orbit around it, is a matter of initial conditions. The fate of such particles can be determined only if we fully decouple the degrees of freedom and integrate the corresponding differential equations. However, the case of $L_z=3.5M$ yields a region enclosed by an ellipse, outside the horizon for which massive particles are trapped there; they cannot escape to infinity, nor fall into the black hole. For counter-rotating trajectories $(L_z<0)$ the structure is qualitatively the same.

\section{Slowly Rotating  Kerr-like Black Holes with Axionic Backreaction}
\label{sec3}

Having reviewed some well known results of the slowly rotating Kerr black hole, we may move now on to the main subject of our work, which is the
study of the angular momentum of a system involving axion fields in CS gravity, coupled to the gravitational CS topological term, and
back-reacting on the rotating black-hole geometry. Such systems
have been most recently studied in \cite{Chatzifotis:2022mob}, extending previous analyses~\cite{kerr1,kerr2,kerr3,yunes1,yunes2}, where the back reaction has been studied perturbatively in an appropriate parameter  for slowly rotating black holes but in a way that allows one to approach arbitrarily close the horizon. The gravitational theory we considered is described by the action
\begin{equation}
	S=\int d^4x \sqrt{-g}\left[\frac{R}{2\kappa^2}-\frac{1}{2}(\partial_\mu b)(\partial^\mu b)-A b 	R_{CS}\right].
	\label{eq:Action}
\end{equation}
The term denoted as $R_{CS}$, the Chern-Simons topological term, is constructed from the contraction of the Riemann tensor to the dual Riemann tensor and reads
\begin{equation}
	\label{RCS}
	R_{CS}= \frac{1}{2}R^{\mu}_{\,\,\,\nu\rho\sigma}\widetilde{R}^{\nu\,\,\,\,\rho\sigma}_{\,\,\,\mu},
\end{equation}
where the symbol $\widetilde{(\dots)}$ denotes the dual of the Riemann tensor, defined as
\begin{equation}
	\label{dualriem}
	\widetilde{R}_{\alpha\beta\gamma\delta}=\frac{1}{2}R_{\alpha\beta}^{\,\,\,\,\,\,\,\,\rho\sigma}\varepsilon_{\rho\sigma\gamma\delta},
\end{equation}
with $\varepsilon_{\rho\sigma\kappa\lambda} = \sqrt{-g(x)} \, \hat \epsilon_{\rho\sigma\kappa\lambda} $  the covariant Levi-Civita  under the convention that the symbol $\hat \epsilon_{0123}=1$, {\it etc}. Regarding the constants of the theory under consideration,  $\kappa=M_p^{-1}$ is the inverse of the reduced Planck mass, while $A$ denotes the strength of the coupling in units of length. The matter content of the theory contains the pseudoscalar field $b$, which denotes the axion.

Variation of the action with respect to the metric $g_{\mu\nu}$ and the axion $b$ yields the following equations of motion
\begin{align}
	\label{grav}
	&G_{\mu\nu}=\kappa^2 T^b_{\mu\nu}+4 \kappa^2 A C_{\mu\nu}~,\\
	\label{Axion}
	&\square b=A R_{CS}~,
\end{align}
where $T^b_{\mu\nu}$ is the usual stress energy tensor associated with the kinetic term of a matter field,
\begin{equation}
	\label{stressaxion}
	T^b_{\mu\nu}=\nabla_\mu b\nabla_\nu b-\frac{1}{2}g_{\mu\nu}(\nabla b)^2,
\end{equation}
while $C_{\mu\nu}$ is the Cotton tensor derived from the metric variation of $b R_{CS}$ and reads
\begin{equation}
	\label{cotton}
	C_{\mu\nu}=-\frac{1}{2}\nabla^{\alpha}\left[(\nabla^{\beta} b) \tilde{R}_{\alpha\mu\beta\nu}+(\nabla^{\beta} b) \tilde{R}_{\alpha\nu\beta\mu}\right].
\end{equation}
The above modification of the gravitational equations of motion and specifically the non-vanishing divergence of the Cotton tensor yields an apparent exchange of energy between the axion and gravity. Specifically, we have that
\begin{equation}
	\label{exchange}
\nabla_\mu C^{\m\n}=-\frac{1}{4}\left(\nabla^\n b\right)R_{CS}
\end{equation}
which implies, through the Bianchi identity, that $T^b_{\m\n}$ itself is not conserved. Indeed, it is the effective energy-momentum tensor given by the r.h.s. of (\ref{grav}) that is conserved, since $\nabla_\mu G^{\m\n}=0$ leads to
\begin{equation}
\nabla_\mu T_b^{\m\n}=A\frac{1}{4}\left(\nabla^\n b\right)R_{CS}
\label{conservation}
\end{equation}
Thus, the presence of the CS anomaly term acts as a source of energy for the axion field, when is non-vanishing, as is the case, when one considers gravitational wave perturbations \cite{Mavromatos:2021urx,Alexander:2004us} or rotating black holes \cite{Chatzifotis:2022mob,kerr1,kerr2,kerr3}, where the latter is of interest in the current study. Moreover, equation (\ref{conservation}) leads to the axionic field equation (\ref{Axion}), as one can easily establish in view of $\nabla_\m T_b^{\m\n}=\square b\nabla^\n b$, which means that this exchange of energy governs the dynamics of the axion. \par

In \cite{Chatzifotis:2022mob}, we extracted a slowly rotating local black hole solution of the CS gravity keeping a first order approximation scheme on the angular momentum per unit mass parameter $a$. We did not consider additional approximation schemes on the coupling. Rather, the line element of the geometry contains all orders of the coupling $A$, which means that we can deduce conclusions for the behavior of the system in the strong coupling regime or to explore the physics near the event horizon.  \par
Our solution reads:
\begin{equation}
	ds^2= -\left( 1-\frac{2M}{r} \right)dt^2+\frac{dr^2}{\left( 1-\frac{2M}{r} \right)}+r^2d\Omega^2+2g_{t\phi}(r,\th)dtd\phi~,
	\label{eq:metricsolution}
\end{equation}
where
\begin{equation}
	g_{t\phi}=-\left(\frac{2 M}{r}+w(r)\right)a\;\sin^2(\th)~,
	\label{tphicomponentsolution}
\end{equation}
and
\begin{equation}
	w(r)=\sum_{n=4}^{\infty}\frac{d_n M^{n-2}}{r^{n-2}}~,
	\label{eqw}
\end{equation}
with the coefficients $d_n$ are given by the recurrence relation
 \begin{equation}
	d_n=\frac{2(n-5)^2(n-1)}{n(n-6)(n-3)}d_{n-1}+\frac{576A^2\k^2}{n(n-3)M^4}d_{n-6},\;\;\text{for}\;n\geq 10~,
	\label{dnequation}
\end{equation}
under the initial values of
\begin{equation}\label{ddef}
	d_4=d_5=0\;\;,\;\;d_6=-5\g^2\;\;,\;\;d_7=-\frac{60\g^2}{7}\;\;,\;\;d_8=-\frac{27\g^2}{2}\;\;,\;\;d_9=0,
\end{equation}
where
\begin{align}\label{gammadef}
\displaystyle \gamma^2=\frac{A^2\k^2}{M^4}\,,
\end{align}
 is a dimensionless parameter, which measures the strength of the backreaction on the geometry. This parameter will play an important r\^ole in our subsequent analysis.

 The axion matter field reads
\begin{equation}
	b(r,\th)=-\frac{r^5 (w/r^2)^\prime}{24\gamma^2 M^5}a A\cos(\th)\sim a A\cos\theta\left(-\frac{5}{4M r^2}-\frac{5}{2r^3}-\frac{9M}{2r^4}\right)+\mathcal{O}(A^m),\,\,\,\,\,\,\text{for}\,\,\, m = 2n + 1, \, \, n \in  \mathbb Z^+,,
	\label{axionsol}
\end{equation}
The function $w(r)$ appearing in (\ref{tphicomponentsolution}) corresponds to the deformation of the Kerr metric due to the specific coupling of the axion field to the CS term; it is the result of the backreaction of the axion hair to the background metric. Since the function is expressed in terms of power series of the dimensionless parameter $\gamma$, the deformation of the Kerr metric depends not only on the coupling constant $A$, but also on the black hole mass. Thus, it is better to understand the solution as a power series of $\gamma$, rather than the coupling constant $A$ itself.

By introducing the axion matter backreaction term, $g_{t\phi}$ is given by (\ref{tphicomponentsolution}) and then, the effective potential of the geodesics can be written as
\begin{equation}
	V_{eff}(r,\th)=V_{eff}^K(r,\th)+\frac{L_z a w(r)}{r^2}~.
	\label{newpot}
\end{equation}
Because of the presence of the backreaction term $w(r)$ in the effective potential (\ref{newpot}) we expect  to see a qualitatively different behavior of particles around the black hole. To explore such a possibility, we plot the graphs of the effective potential and CZV for increasing $\gamma$. We plot in Figs. \ref{timelikeeffectiveKerr1} - \ref{timelikeeffectiveKerr2} the effective potential  for increasing $\gamma$ for positive and negative angular momentum. We observe a completely new structure for $\gamma=2$. The effective potential becomes repulsive for counter-rotating orbits close to the horizon and a new local minimum appears, signifying for a new family of bound orbits. Moreover, this repulsive nature of the effective potential, beyond the new family of bound orbits, signifies scattering of particles that arrive too close to the event horizon. Thus, counter-rotating orbits are prevented to fall into the black hole. We stress that the derivation of the following figures for $\gamma\geq1$ is only possible if one considers (in principle) all of the correction terms of the backreaction metric component or at least enough correction terms, such that the possible margin of error is miniscule. We note that this is possible, because the power series of (\Ref{eqw} converges, as was proven in \cite{Chatzifotis:2022mob}.}\pagebreak
\begin{figure}[h!]
	\centering
	\includegraphics[width=0.5\textwidth]{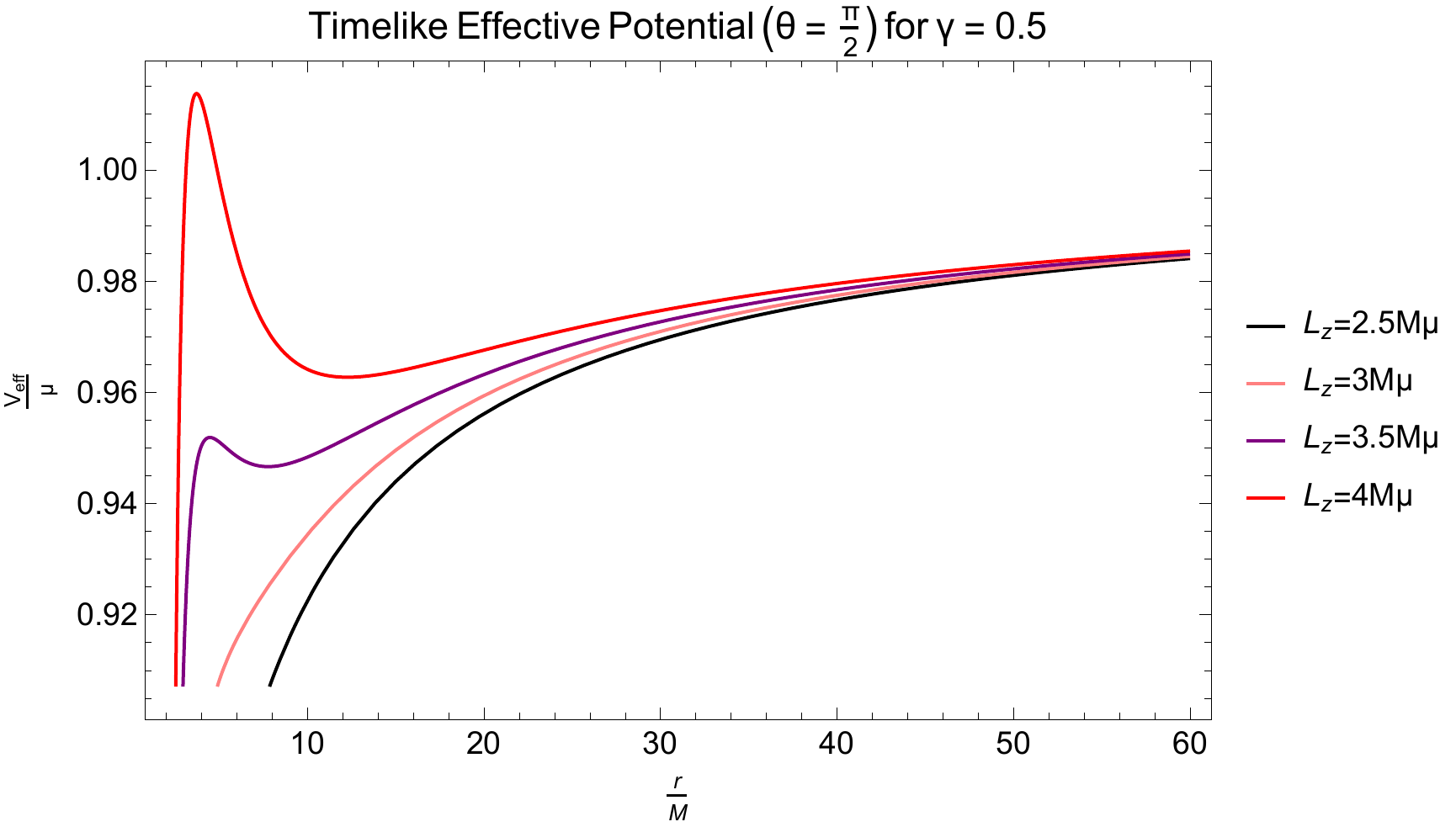}\hfil\includegraphics[width=0.5\textwidth]{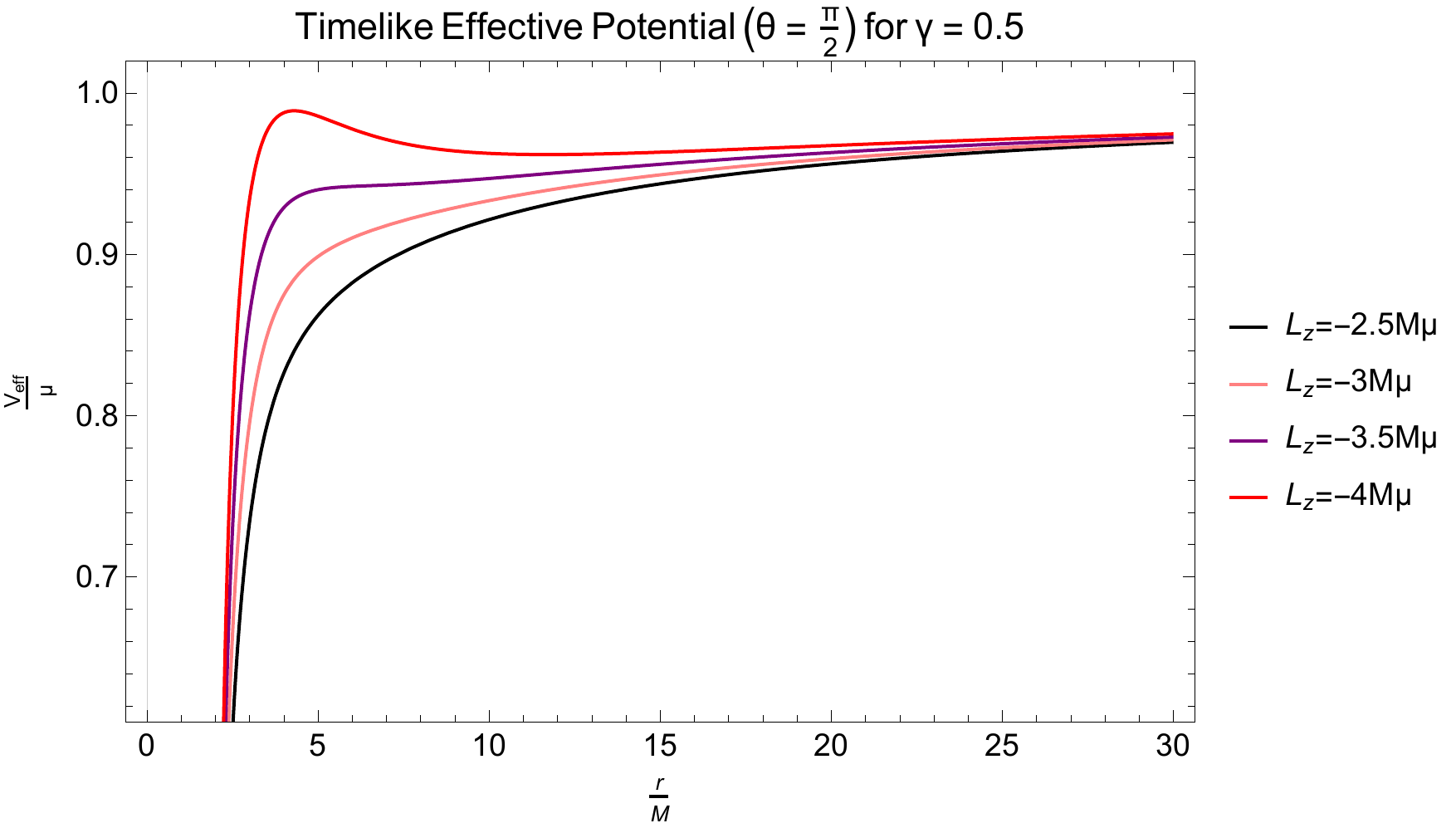}
	\caption{Effective potential at the equatorial plane for timelike geodesics for the slowly rotating deformed Kerr metric. ($\gamma=0.5,\; a=0.1$) }
	\label{timelikeeffectiveKerr1}
\end{figure}

\begin{figure}[h!]
	\centering
	\includegraphics[width=0.5\textwidth]{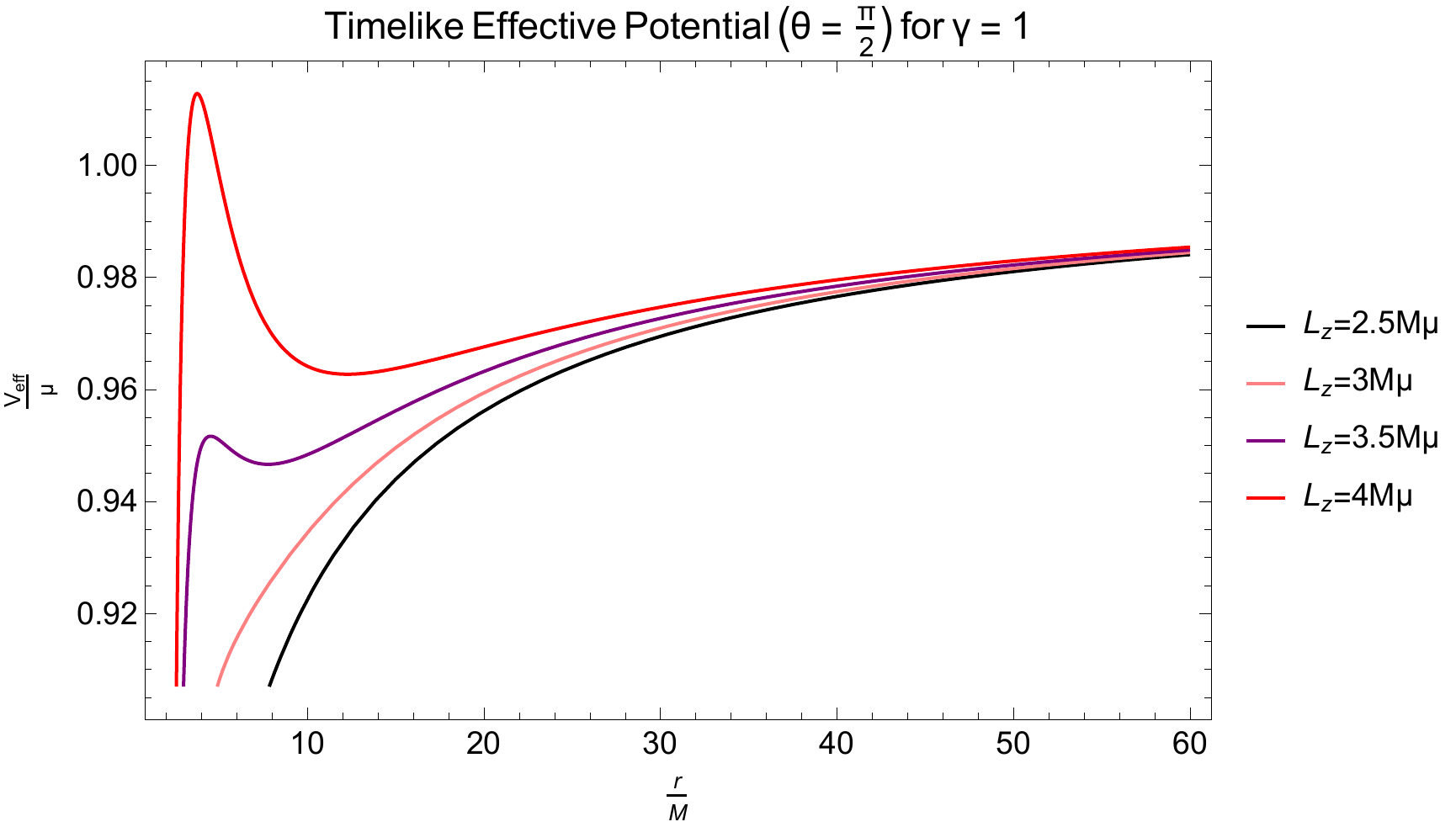}\hfil\includegraphics[width=0.5\textwidth]{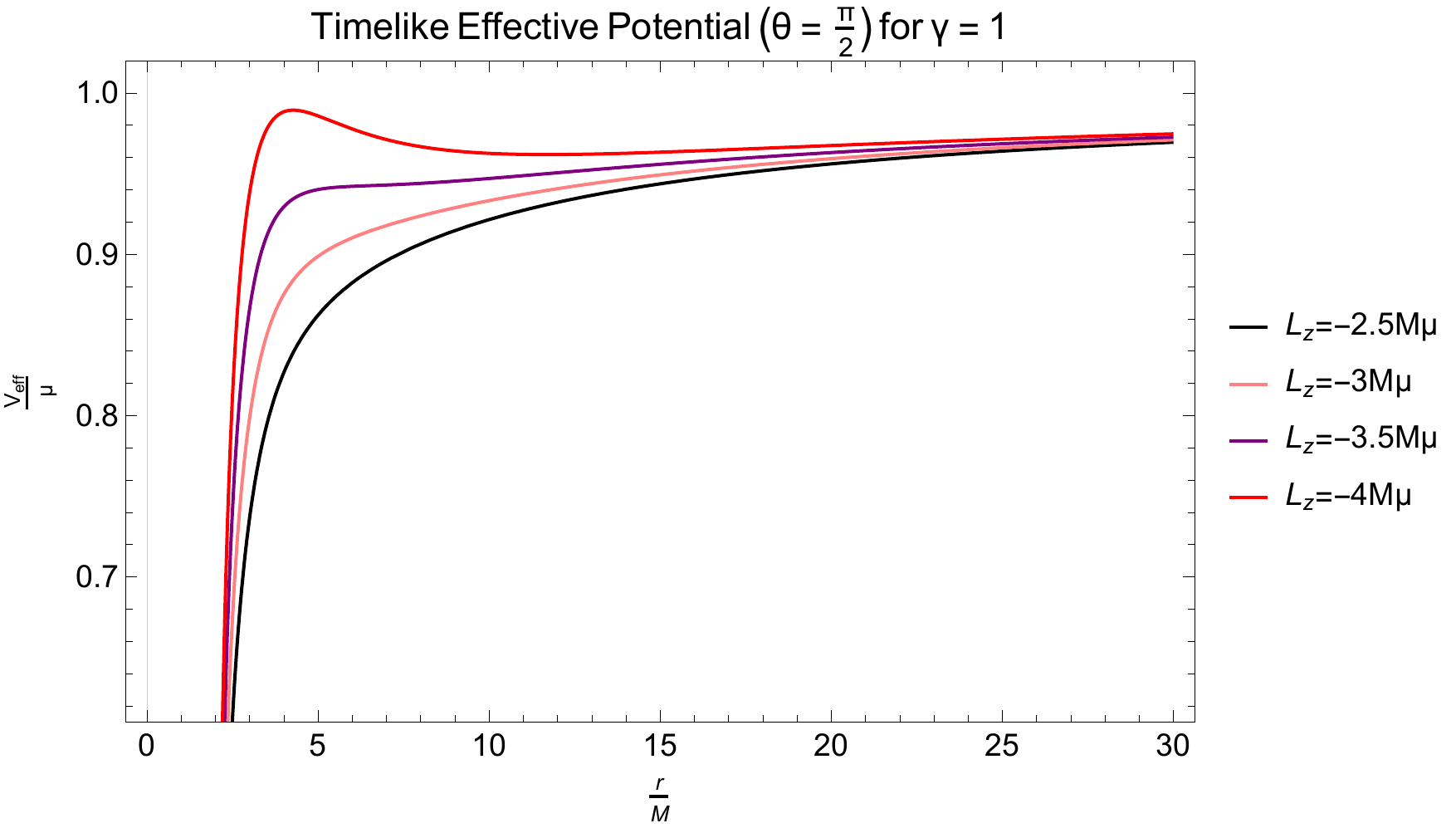}
	\caption{Effective potential at the equatorial plane for timelike geodesics for the slowly rotating deformed Kerr metric. ($\gamma=1,\; a=0.1$) }
	\label{timelikeeffectiveKerr}
\end{figure}

\begin{figure}[h!]
	\centering
	\includegraphics[width=0.5\textwidth]{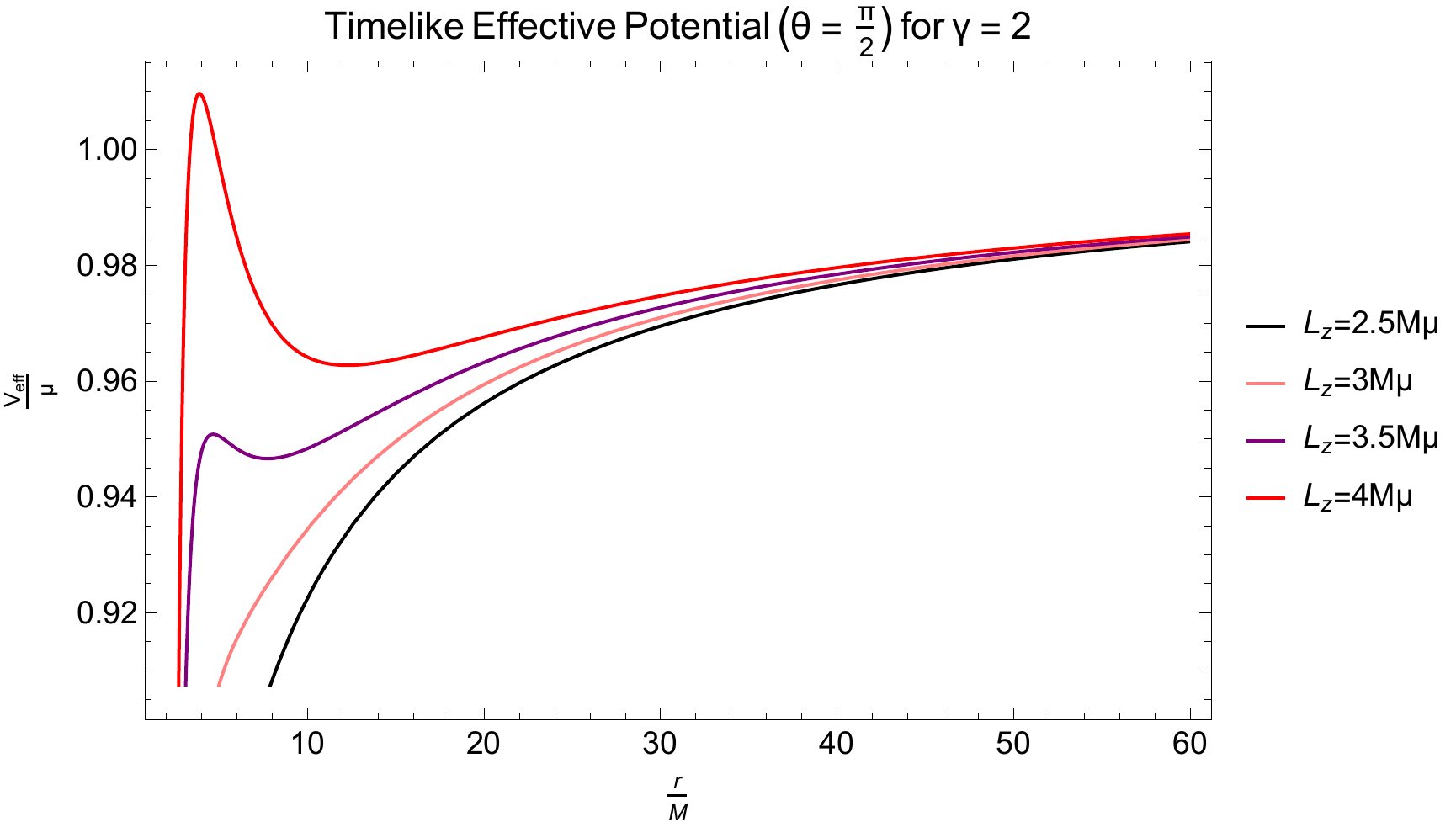}\hfil\includegraphics[width=0.5\textwidth]{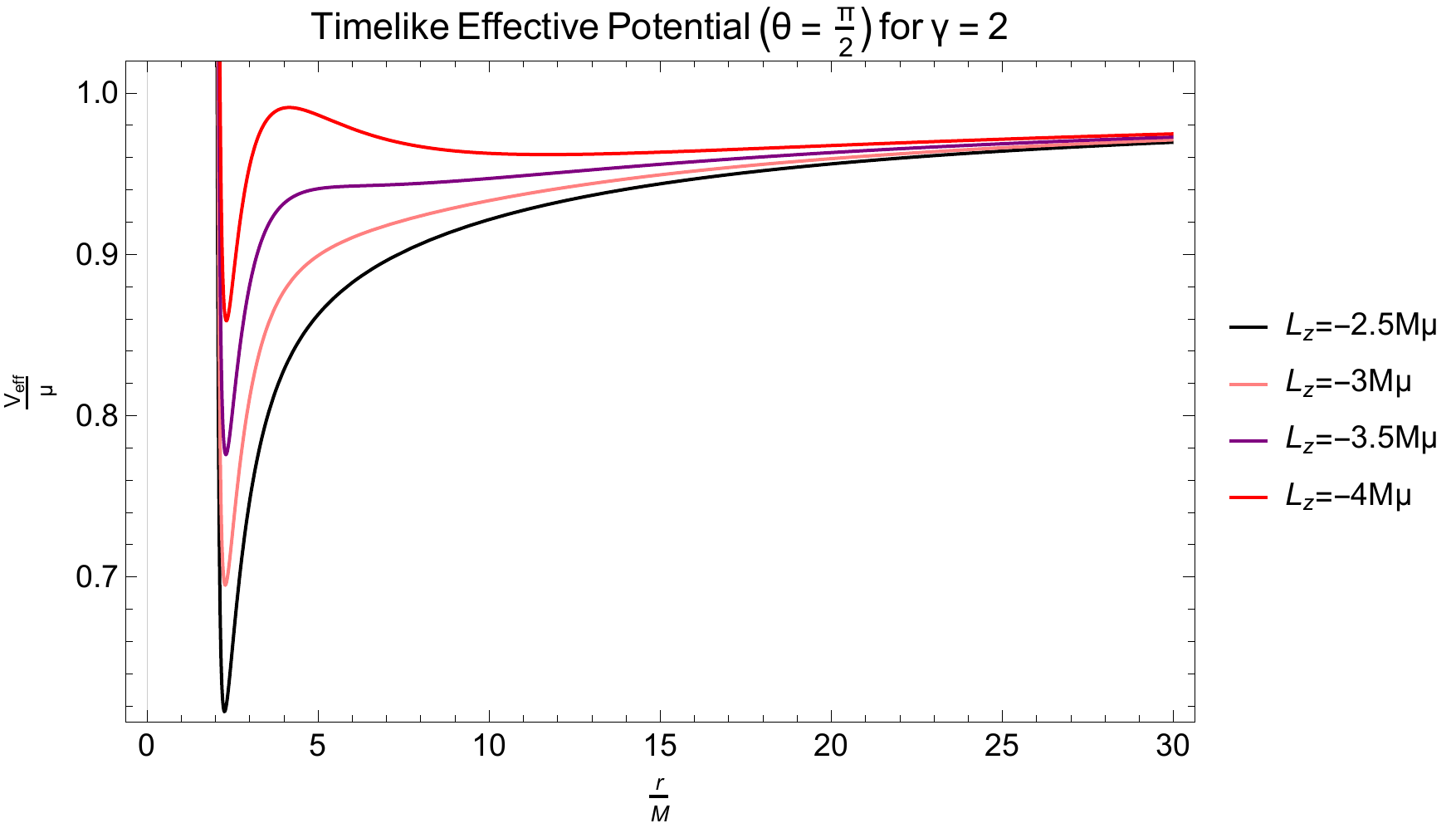}
	\caption{Effective potential at the equatorial plane for timelike geodesics for the slowly rotating deformed Kerr metric. ($\gamma=2,\; a=0.1$) }
	\label{timelikeeffectiveKerr2}
\end{figure}

\begin{figure}[h!]
	\centering
	\includegraphics[width=0.40\textwidth]{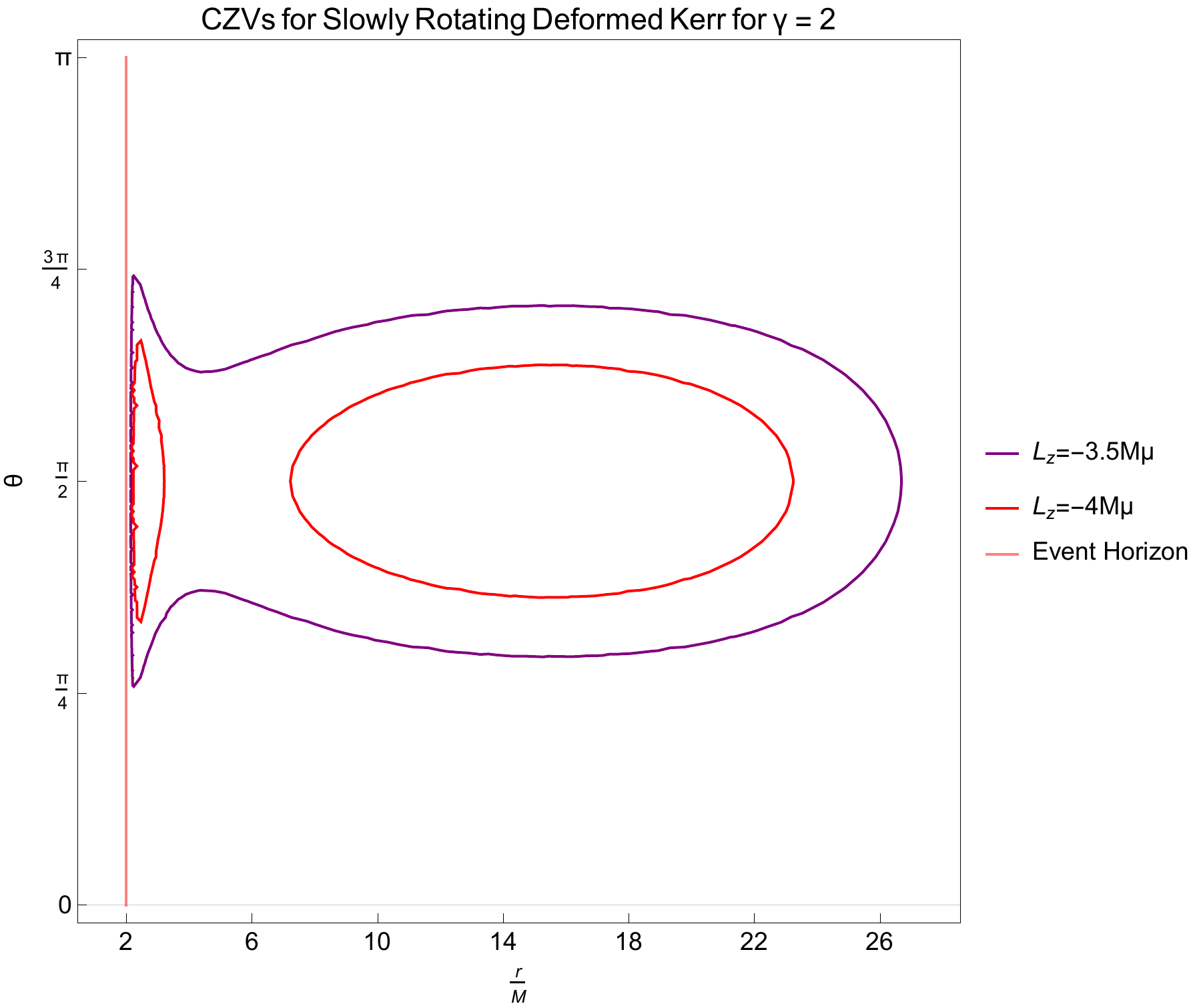}
	\caption{Effective potential at the equatorial plane for timelike geodesics for the slowly rotating deformed Kerr metric. ($\gamma=2,\; a=0.1$) }
	\label{CZVsdeformedKerr}
\end{figure}

This new structure may also be verified from the following CZV figure, where turning points close to the horizon appear. In the CZV of Fig.  \ref{CZVsdeformedKerr}, we see a similar structure with that of the slowly rotating Kerr far away from the horizon, but turning points close to the horizon appear, preventing particles to fall into the black hole.
In \cite{Pugliese:2011xn} the effective potential of the Kerr metric has been studied extensively, for the black hole, the extreme black hole and the naked singularity case. The extreme case of the Kerr metric corresponds to a highly rotating black hole and since there is no matter coupled to gravity, corresponds to a highly rotating event horizon. In the extreme case, the effective potential has a similar structure as that in our case for $\gamma=2$. To be more precise, the aforementioned repulsive nature appears also in the extreme case of the Kerr metric for the counter-rotating orbits, which are distinguished from the co-rotating ones in a similar fashion. Thus, this effect seems to be enforced by a highly rotating region, meaning that a highly rotating event horizon is responsible for this behavior of the effective potential. This is actually true even in our case, where the whole spacetime remains slowly rotating but its internal structure is comprised of highly rotating systems; the axion and gravitational field, non-minimally coupled with each other. In the next section, we study the angular momentum of the spacetime and its contributions coming from the event horizon and the axion matter field, in order to make clear how such a behavior is possible.

\section{The Angular Momentum of the Axionic Black Hole}
\label{sec4}

Following the standard formalism of \cite{bardeen_carter}, we wish to analytically compute the angular momentum of the event horizon of our black hole solution.
We recall that the Killing vector corresponding to polar isometry of the spacetime, which reads $\xi=\partial_\phi$, obeys the identity
\begin{equation}
	\label{4.1}
\nabla_\b\nabla^\b\xi^\a=-R^\a\!_\b \xi^\b~,
\end{equation}
where $R^\a_\b$ is the Ricci tensor. We can integrate the above equation over a hypersurface $S$, yielding
\begin{equation}
	\label{4.2}
\int_S d\Sigma_\a \nabla_\b\nabla^\b\xi^\a=-\int_S d\Sigma_\a R^\a\!_\b \xi^\b~.
\end{equation}
The tensor $\nabla^\a \xi^\b$, in view of the Killing equation, is antisymmetric, which means that we can transform the left handed integral to an integral over a 2-surface, i.e. the boundary $\partial S$ of $S$
\begin{equation}
	\label{4.3}
\int_{\partial S}d\Sigma_{\a\b}\nabla^\a \xi^\b=-\int_S d\Sigma_\a R^\a\!_\b \xi^\b~.
\end{equation}
We take the hypersurface $S$ to be the exterior of the black hole and as such the boundary consists of a 2-surface at spatial infinity $\partial S_\infty$ and the event horizon $\mathcal{H}$. Therefore, the left hand side of the above equation yields
\begin{equation}
	\label{4.4}
	\int_{\partial S}d\Sigma_{\a\b}\nabla^\a \xi^\b=\int_{\partial S_{\infty}}d\Sigma_{\a\b}\nabla^\a \xi^\b+\int_{\mathcal{H}}d\Sigma_{\a\b}\nabla^\a \xi^\b~.
\end{equation}
The integral at spatial infinity corresponds to Komar's integral, which gives us the total angular momentum $J$ as measured at spatial infinity. Thus, we solve (\Ref{4.3}) for $J$ and find
\begin{equation}
	\label{4.5}
	J=-\frac{1}{8\pi}\int_{\mathcal{H}}d\Sigma_{\a\b}\nabla^\a\xi^\b-\frac{1}{8\pi}\int_S d\Sigma_\a R^\a\!_\b \xi^\b~.
\end{equation}
Furthermore, according to the gravitational equations of motion (\Ref{grav}), the Ricci tensor can be written in terms of the energy-momentum tensor, yielding
\begin{equation}
	\label{4.6}
	J=-\frac{1}{8\pi}\int_{\mathcal{H}}d\Sigma_{\a\b}\nabla^\a\xi^\a-\int_S d\Sigma_\a \left( T^\a\!_\b-\frac{1}{2}\delta^\a\!_\b T   \right) \xi^\b~.
\end{equation}
The above terms can be interpreted as the two contributions to the total angular momentum as measured at infinity. The first one is interpreted as the black hole's angular momentum,
\begin{equation}
	\label{4.7}
J_H=-\frac{1}{8\pi}\int_{\mathcal{H}}d\Sigma_{\a\b}\nabla^\a\xi^\a~,
\end{equation}
 while the second one, can be interpreted as the angular momentum of the matter field outside the event horizon:
 \begin{equation}
 	\label{4.8}
J_M=-\int_S d\Sigma_\a \left( T^\a\!_\b-\frac{1}{2}\delta^\a\!_\b T   \right) \xi^\b~.
 \end{equation}
Obviously, if there is no matter ($T_{\m\n}=0$) the only contribution comes from the black hole angular momentum and $J=J_H$; this is the case of the Kerr solution, which is a vacuum solution. \par
In order to calculate the black hole's angular momentum we have to firstly define the surface element $d\Sigma_{\a\b}$. We may take the two surfaces that construct $\partial S$ to be those of $t=const$ and $r=const$ and at the end of the calculations we can take the appropriate limits\footnote{We avoid the use of a single null normal vector and the corresponding auxiliary one to simplify calculations.}.  Then, we get the two normal one-forms $ n_\m=n_o\partial_\m t$ and $ \sigma_\m=\sigma_o\partial_\m r$, where $n_0$ and $\sigma_0$ are normalization constants. For the latter, the normalization is chosen such that $n_\m n^\m=-\sigma_\m\sigma_\m=-1$. Up to $\mathcal{O}(a)$, these relations give us $n_o=1/\sigma_0=\pm\sqrt{1-2M/r}$. In order to have a future and outgoing directed orientation of the 2-surfaces we choose $n_0<0$ and $\sigma_0>0$, i.e.
 \begin{equation}
 	\label{4.9}
\begin{aligned}
& n_\m=-\sqrt{1-\frac{2M}{r}}\partial_t~,\\
& \sigma_\m=\frac{1}{\sqrt{1-\frac{2M}{r}}}\partial_r~.
\end{aligned}
 \end{equation}
In terms of these vectors the surface element reads: $d\Sigma_{\a\b}=\sqrt{g^{(2)}}n_\a\sigma_\b d^2x$, where $g^{(2)}$ is the determinant of the two dimensional induced metric,  $\sqrt{g^{(2)}}=r^2\sin\th$. Moreover, $\nabla^\a\xi^\b=g^{\a\m}\Gamma^\b_{\m\phi}$, which finally gives us
\begin{equation}
	\label{4.10}
J_H=-\frac{1}{8\pi}\int_0^\pi d\th\int_0^{2\pi}d\phi r^2\sin\th\; n_\m\sigma_\n g^{\m\a}\Gamma^\n_{\a\phi}~.
\end{equation}
Then in our case, with backreaction of matter to the Kerr-like black hole the calculation gives
\[
n_\m\sigma_\n g^{\m\a}\Gamma^\n_{\a\phi}=\frac{\sin^2(\th)(-6M-2rw(r)+r^2w^\prime(r))a}{2r^2}~.
\]
By substitution and integration, we finally get
\begin{equation}
	J= \left[1+\frac{2rw(r)-r^2w^\prime(r)}{6M}  \right]Ma~.
	\label{bhangularmomentumtotal}
\end{equation}
Taking the limit $r\rightarrow +\infty$, we get the angular momentum as given by the Komar's integral, which obviously gives as the total angular momentum equal to $J=Ma$, as $w$ and $w^\prime$ fall faster than $r^2$. Taking $r=2M$, we find the black hole's angular momentum
\begin{equation}
J_H=\left[ 1+\frac{2rw(r)-r^2w^\prime(r)}{6M}  \right]_{r=2M}Ma~.
\label{bhangularmomentum}
\end{equation}
For the axionic matter, it is straightforward to obtain:
\begin{equation}
	\label{matteram}
J_M=-\int d^3x n_0\sqrt{g^{(3)}} T^t\!_\phi~,
\end{equation}
where $g^{(3)}$  is the determinant of the three dimensional induced metric of the hypersurface $t=const$ and reads $\sqrt{g^{(3)}}=r^2\sin\th/\sqrt{1-2M/r}$. Thus,
\begin{equation}
	\label{matteram1}
J_M=2\pi\int_0^\pi d\th \sin\th\int_{2M}^{+\infty}r^2T^t\!_\phi~.
\end{equation}
Substituting $T^t\!_\phi$ up tp $\mathcal{O}(a)$ and performing the angular integrations, we get
\begin{equation}
	\label{matteram2}
J_M=\frac{a}{6}\int^{+\infty}_{2M}\left[2w(r)-r^2w^{\prime\prime}(r)\right]dr~.
\end{equation}
The integrand can be expressed as a total derivative \[ 2w(r)-r^2w^{\prime\prime}(r)=-\frac{d}{dr}\left(r^2w(r)-2rw(r)\right) \]~,
and since $r^2w^\prime,rw(r)\rightarrow0$, as $r\rightarrow +\infty$, we get the final result
\begin{equation}
J_M=-\left[  \frac{2rw(r)-r^2w^\prime(r)}{6M}   \right]_{r=2M}Ma~.
\label{matterangularmomentum}
\end{equation}
Thus, (\ref{bhangularmomentum}) and (\ref{matterangularmomentum}) give for the total angular momentum
\begin{equation}
	\label{totang}
J=J_H+J_M=Ma~.
\end{equation}

Thence, the total angular momentum is equal to $Ma$ as in the Kerr case, but in this case there is a non-trivial internal structure consisted of the black hole and axion angular momentum. The effect of the axion is to rotate the black hole, in the sense that it is necessary for the spacetime to be rotating in order for the axion to be dynamical.\footnote{For a spherically symmetric background $\mathcal{R}_{CS}=0$ and the axion is non-dynamical.} In addition, as previously mentioned, the parameter $\gamma$, which appears in the correction function $w(r)$ is a measure of how strong the backreaction on the geometry is. Thus, as the backreaction gets stronger, i.e. for increasing $\gamma$, we expect for black hole rotation to change. The axion cloud around the black hole acquires an angular momentum in such a way that the total angular momentum remains constant and equal to $Ma$. As $\gamma$ increases the black hole angular momentum decreases until a critical value, $\gamma_c$, for which the black hole has no angular momentum at all. After that value, the black hole starts to rotate in the opposite direction with increasing magnitude. Hence, the magnitude of each particular angular momentum can reach large values, but the angular momentum of the spacetime as a whole, as measured by an observer at infinity, remains as if the background were slowly rotating.

\begin{figure}[h!]
	\centering
	\includegraphics[width=0.5\textwidth]{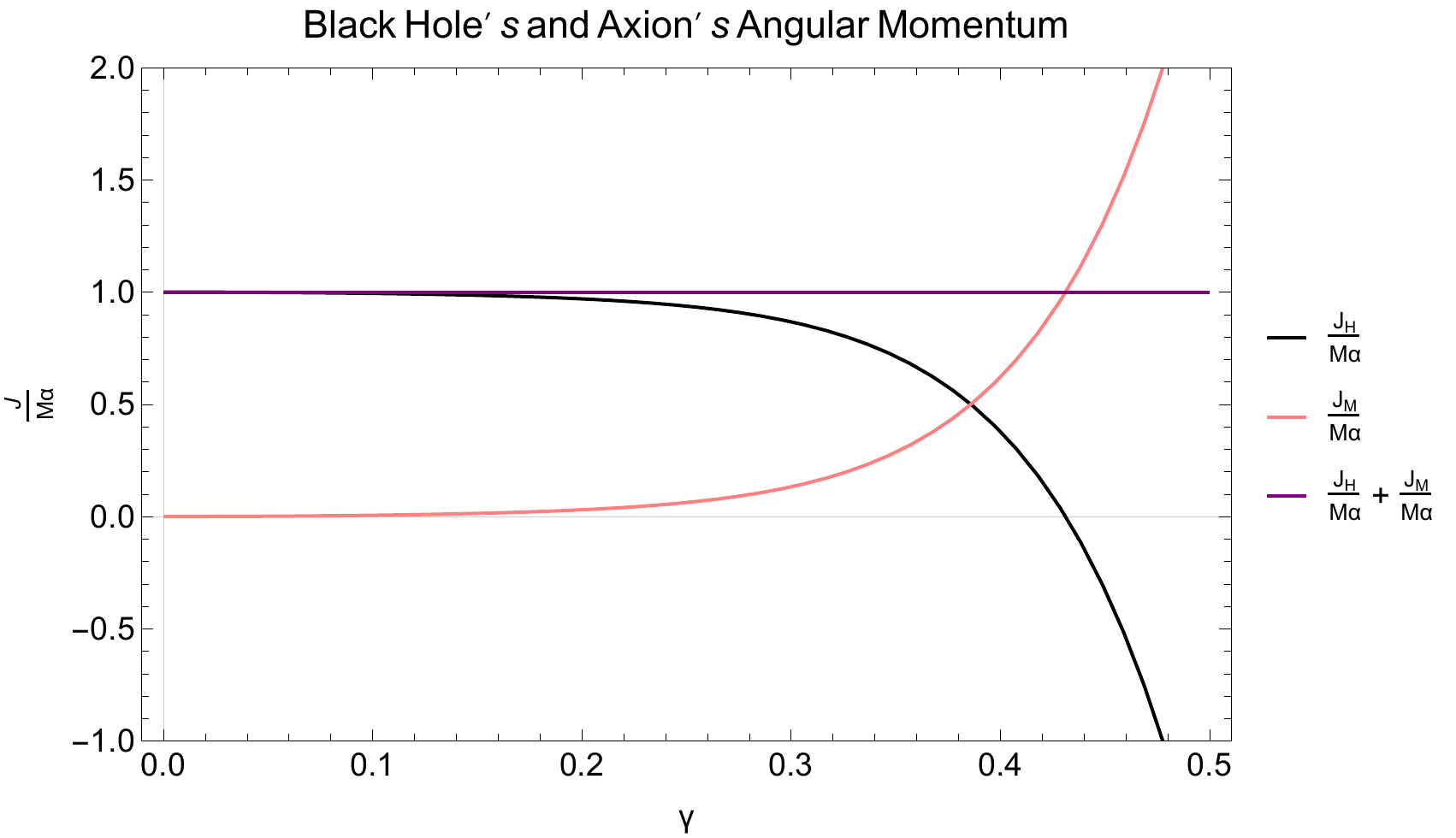}
	\caption{The total angular momentum of the system of the Kerr-like black hole and the surrounding axionic matter.  }
	\label{fig7}
\end{figure}

One may understand this, by having in
mind two competitive systems; the vacuum spacetime (vacuum solution of the Einstein’s equations) and the
axion, the pseudoscalar field which couples to the CS term. If we turn off the coupling, the system reduces to the Kerr metric which is a solution of GR in vacuum. This solution describes a rotating Kerr black hole. Turning on the coupling, the axion becomes dynamical, acquires angular momentum and tends to rotate the black hole in the other way. When the coupling is strong enough, the spacetime around the black hole is counter rotating with regard to the measured Komar integral of the whole spacetime, as can be seen in Fig.~\ref{fig7}. This is a manifestation of the exchange of energy between the axionic matter and  the gravitational field.

\section{Conclusions and Outlook} \label{sec5}

In this work we have studied a system consisting of a pseudoscalar (axion-like) field coupled to a Chern-Simons gravitational anomalous term in a Kerr-like black hole space time, which was generated by the backreaction of the axion field to a (slowly) rotating Kerr black hole background. The system is characterised by an appropriate dimensionless coupling constant, $\gamma$, proportional to the ratio of the axion-Chern-Simons coupling over the square of the black hole mass. This parameter measures the strength of the backreaction of the axion field on the spacetime geometry.

We carried out an analysis of the angular momentum of this system. The axion matter outside the horizon of the black hole acquires an angular momentum in such a way that the total angular momentum of the system of the modified Kerr black hole and the axionic matter remains constant. We found that, as $\gamma$ increases,  the black hole angular momentum decreases and there exists a critical value $\gamma_c$ beyond which the black hole angular momentum increases, with the black hole rotating in the opposite direction. This effect can be understood as a counterplay of two competing systems, the Kerr-like black hole and the axionic matter rotating outside the horizon of the black hole. As the coupling of the axionic matter to gravity is getting stronger, there is an increasing exchange of energy between the axionic matter and the gravitational field expressed by the Chern-Simons  term, which, when the coupling exceeds a critical value, results in the aforementioned phenomenon of reversal of the angular momentum of the rotating black hole.

It would be interesting to study the stability of the modified Kerr black hole with the axionic matter outside the horizon. We expect that, as $\gamma$ is increases, the system will develop an instability. In addition, since our study was restricted to a slowly-rotating black hole, it would be interesting to extend this work to fast rotating black holes, at least to order $a^{2}$ in the angular parameter,  and see if the counter rotating behaviour of the Kerr-like black hole still persists.

\acknowledgments

We thank Kostas Kokkotas for valuable discussions.
The work of N.C. and E.P is supported by the research project of the National Technical University of Athens (NTUA)
65232600-ACT-MTG:  {\it Alleviating Cosmological Tensions Through Modified Theories of Gravity}; that of P.D. is partially supported by
a NTUA scholarship, while the work  of N.E.M. is supported in part by
the UK Science and Technology Facilities  research Council (STFC) under the research grant ST/T000759/1.
N.E.M.  also acknowledges participation in the COST Association Action CA18108 ``{\it Quantum Gravity Phenomenology in the Multimessenger Approach (QG-MM)}''.


\begin{thebibliography}{99}

%\cite{Kerr:1963ud}
\bibitem{kerr}
R.~P.~Kerr,
``Gravitational field of a spinning mass as an example of algebraically special metrics,''
Phys. Rev. Lett. \textbf{11}, 237-238 (1963)
%di:10.1103/PhysRevLett.11.237
%1907 citations counted in INSPIRE as of 21 Jun 2022



%\cite{Jackiw:2003pm}
\bibitem{Jackiw}
R.~Jackiw and S.~Y.~Pi,
``Chern-Simons modification of general relativity,''
Phys. Rev. D \textbf{68}, 104012 (2003)
%doi:10.1103/PhysRevD.68.104012
[arXiv:gr-qc/0308071 [gr-qc]].
%527 citations counted in INSPIRE as of 15 Jun 2022


%\cite{Zanelli:2005sa}
\bibitem{Zanelli:2005sa}
J.~Zanelli,
``Lecture notes on Chern-Simons (super-)gravities. Second edition (February 2008),''
[arXiv:hep-th/0502193 [hep-th]].
%210 citations counted in INSPIRE as of 20 Jan 2022

%\cite{Alexander:2009tp}
\bibitem{Alexander}
S.~Alexander and N.~Yunes,
``Chern-Simons Modified General Relativity,''
Phys. Rept. \textbf{480}, 1-55 (2009)
%doi:10.1016/j.physrep.2009.07.002
[arXiv:0907.2562 [hep-th]].
%454 citations counted in INSPIRE as of 16 Jun 2022



\bibitem{yunes1}
N.~Yunes and F.~Pretorius,
``Dynamical Chern-Simons Modified Gravity. I. Spinning Black Holes in the Slow-Rotation Approximation,''
Phys. Rev. D \textbf{79}, 084043 (2009)
%doi:10.1103/PhysRevD.79.084043
[arXiv:0902.4669 [gr-qc]].
%221 citations counted in INSPIRE as of 20 Dec 2021

%\cite{Yagi:2012ya}
\bibitem{yunes2}
K.~Yagi, N.~Yunes and T.~Tanaka,
``Slowly Rotating Black Holes in Dynamical Chern-Simons Gravity: Deformation Quadratic in the Spin,''
Phys. Rev. D \textbf{86}, 044037 (2012)
[erratum: Phys. Rev. D \textbf{89}, 049902 (2014)]
%doi:10.1103/PhysRevD.86.044037
[arXiv:1206.6130 [gr-qc]].
%144 citations counted in INSPIRE as of 20 Dec 2021
%\cite{Mavromatos:2021urx}



\bibitem{kerr1}
 B.~A.~Campbell, M.~J.~Duncan, N.~Kaloper and K.~A.~Olive,
``Axion hair for Kerr black holes,''
Phys. Lett. B \textbf{251}, 34-38 (1990)
%doi:10.1016/0370-2693(90)90227-W
%89 citations counted in INSPIRE as of 19 Dec 2021

\bibitem{kerr2}
 B.~A.~Campbell, M.~J.~Duncan, N.~Kaloper and K.~A.~Olive,
``Gravitational dynamics with Lorentz Chern-Simons terms,''
Nucl. Phys. B \textbf{351}, 778-792 (1991)
%doi:10.1016/S0550-3213(05)80045-8
%88 citations counted in INSPIRE as of 19 Dec 2021


\bibitem{kerr3}
 M.~J.~Duncan, N.~Kaloper and K.~A.~Olive,
``Axion hair and dynamical torsion from anomalies,''
Nucl. Phys. B \textbf{387}, 215-235 (1992)
%doi:10.1016/0550-3213(92)90052-D
%52 citations counted in INSPIRE as of 19 Dec 2021


%\cite{Svrcek:2006yi}
\bibitem{witten}
P.~Svrcek and E.~Witten,
``Axions In String Theory,''
JHEP \textbf{06}, 051 (2006)
%doi:10.1088/1126-6708/2006/06/051
[arXiv:hep-th/0605206 [hep-th]].
%1112 citations counted in INSPIRE as of 16 Jun 2022



\bibitem{CSblackhole}
D. G. Boulware and S. Deser, Phys. Rev. Lett. 55 (1985) 2656;
J. T. Wheeler, Nucl. Phys. B 268 (1986) 737.

%\cite{Gonzalez:2010vv}
\bibitem{Gonzalez:2010vv}
P.~Gonzalez, E.~Papantonopoulos and J.~Saavedra,
``Chern-Simons black holes: scalar perturbations, mass and area spectrum and greybody factors,''
JHEP \textbf{08}, 050 (2010)
%doi:10.1007/JHEP08(2010)050
[arXiv:1003.1381 [hep-th]].
%58 citations counted in INSPIRE as of 20 Jan 2022



%\cite{Sopuerta:2009iy}
\bibitem{Sopuerta:2009iy}
C.~F.~Sopuerta and N.~Yunes,
``Extreme and Intermediate-Mass Ratio Inspirals in Dynamical Chern-Simons Modified Gravity,''
Phys. Rev. D \textbf{80}, 064006 (2009)
%doi:10.1103/PhysRevD.80.064006
[arXiv:0904.4501 [gr-qc]].
%132 citations counted in INSPIRE as of 27 May 2022

%\cite{Canizares:2012is}
\bibitem{Canizares:2012is}
P.~Canizares, J.~R.~Gair and C.~F.~Sopuerta,
``Testing Chern-Simons Modified Gravity with Gravitational-Wave Detections of Extreme-Mass-Ratio Binaries,''
Phys. Rev. D \textbf{86}, 044010 (2012)
%doi:10.1103/PhysRevD.86.044010
[arXiv:1205.1253 [gr-qc]].
%54 citations counted in INSPIRE as of 30 May 2022




%\cite{Chatzifotis:2022mob}
\bibitem{Chatzifotis:2022mob}
N.~Chatzifotis, P.~Dorlis, N.~E.~Mavromatos and E.~Papantonopoulos,
``Scalarization of Chern-Simons-Kerr black hole solutions and wormholes,''
Phys. Rev. D \textbf{105}, no.8, 084051 (2022)
%doi:10.1103/PhysRevD.105.084051
[arXiv:2202.03496 [gr-qc]].
%0 citations counted in INSPIRE as of 14 May 2022

\bibitem{Myung:2020etf}
Y.~S.~Myung and D.~C.~Zou,
%``Onset of rotating scalarized black holes in Einstein-Chern-Simons-Scalar theory,''
Phys. Lett. B \textbf{814}, 136081 (2021)
%doi:10.1016/j.physletb.2021.136081
[arXiv:2012.02375 [gr-qc]].

\bibitem{Myung:2021fzo}
Y.~S.~Myung and D.~C.~Zou,
%``Slowly rotating black holes and their scalarization,''
Phys. Rev. D \textbf{104}, no.6, 064015 (2021)
%doi:10.1103/PhysRevD.104.064015
[arXiv:2103.06449 [gr-qc]].

\bibitem{Zou:2021ybk}
D.~C.~Zou and Y.~S.~Myung,
%``Rotating scalarized black holes in scalar couplings to two topological terms,''
Phys. Lett. B \textbf{820}, 136545 (2021)
%doi:10.1016/j.physletb.2021.136545
[arXiv:2104.06583 [gr-qc]].


%\cite{Bardeen:1972fi}
\bibitem{Bardeen:1972fi}
J.~M.~Bardeen, W.~H.~Press and S.~A.~Teukolsky,
``Rotating black holes: Locally nonrotating frames, energy extraction, and scalar synchrotron radiation,''
Astrophys. J. \textbf{178}, 347 (1972)
%doi:10.1086/151796
%1503 citations counted in INSPIRE as of 30 Mar 2022

%\cite{Lukes-Gerakopoulos:2010ipp}
\bibitem{Lukes-Gerakopoulos:2010ipp}
G.~Lukes-Gerakopoulos, T.~A.~Apostolatos and G.~Contopoulos,
``Observable signature of a background deviating from the Kerr metric,''
Phys. Rev. D \textbf{81}, 124005 (2010)
%doi:10.1103/PhysRevD.81.124005
[arXiv:1003.3120 [gr-qc]].
%61 citations counted in INSPIRE as of 26 May 2022

%\cite{Zelenka:2017aqn}
\bibitem{Zelenka:2017aqn}
O.~Zelenka and G.~Lukes-Gerakopoulos,
``Chaotic motion in the Johannsen-Psaltis spacetime,''
[arXiv:1711.02442 [gr-qc]].
%7 citations counted in INSPIRE as of 26 May 2022

%\cite{Destounis:2021mqv}
\bibitem{Destounis:2021mqv}
K.~Destounis, A.~G.~Suvorov and K.~D.~Kokkotas,
``Gravitational-wave glitches in chaotic extreme-mass-ratio inspirals,''
Phys. Rev. Lett. \textbf{126}, no.14, 141102 (2021)
%doi:10.1103/PhysRevLett.126.141102
[arXiv:2103.05643 [gr-qc]].
%11 citations counted in INSPIRE as of 27 May 2022

%\cite{Destounis:2020kss}
\bibitem{Destounis:2020kss}
K.~Destounis, A.~G.~Suvorov and K.~D.~Kokkotas,
``Testing spacetime symmetry through gravitational waves from extreme-mass-ratio inspirals,''
Phys. Rev. D \textbf{102}, no.6, 064041 (2020)
%doi:10.1103/PhysRevD.102.064041
[arXiv:2009.00028 [gr-qc]].
%11 citations counted in INSPIRE as of 27 May 2022

%\cite{Gonzalez:2018lfs}
\bibitem{Gonzalez:2018lfs}
P.~A.~Gonz\'alez, M.~Olivares, E.~Papantonopoulos and Y.~V\'asquez,
``Motion and collision of particles in a rotating linear dilaton black hole,''
Phys. Rev. D \textbf{97}, no.6, 064034 (2018)
%doi:10.1103/PhysRevD.97.064034
[arXiv:1802.01760 [gr-qc]].
%9 citations counted in INSPIRE as of 04 Jun 2022



\bibitem{carter}
Brandon Carter,
``Global structure of the Kerr family of gravitational fields,''
Phys. Rev. D \textbf{174}, 1559-1571 (1968)


\bibitem{Mavromatos:2021urx}
N.~E.~Mavromatos and J.~Sol\`a Peracaula,
``Inflationary physics and trans-Planckian conjecture in the stringy running vacuum model: from the phantom vacuum to the true vacuum,''
Eur. Phys. J. Plus \textbf{136}, no.11, 1152 (2021)
%doi:10.1140/epjp/s13360-021-02149-6
[arXiv:2105.02659 [hep-th]].
%14 citations counted in INSPIRE as of 06 Jun 202

%\cite{Pugliese:2011xn}
\bibitem{Pugliese:2011xn}
D.~Pugliese, H.~Quevedo and R.~Ruffini,
``Equatorial circular motion in Kerr spacetime,''
Phys. Rev. D \textbf{84}, 044030 (2011)
%doi:10.1103/PhysRevD.84.044030
[arXiv:1105.2959 [gr-qc]].
%96 citations counted in INSPIRE as of 06 Jun 2022

%\cite{Alexander:2004us}
\bibitem{Alexander:2004us}
S.~H.~S.~Alexander, M.~E.~Peskin and M.~M.~Sheikh-Jabbari,
``Leptogenesis from gravity waves in models of inflation,''
Phys. Rev. Lett. \textbf{96}, 081301 (2006)
%doi:10.1103/PhysRevLett.96.081301
[arXiv:hep-th/0403069 [hep-th]].
%227 citations counted in INSPIRE as of 07 Jun 2022

\bibitem{bardeen_carter}
J. M. Bardeen, B. Carter S.W. Hawking,
``The Four laws of black hole mechanics,''
Commun. math. Phys. 31, 161-170 (t973)



\end{thebibliography}
\end{document}